\documentclass[journal=jacsat,manuscript=article]{achemso}
\SectionNumbersOn
\usepackage{chemformula} 
\usepackage[T1]{fontenc} 
\usepackage{overpic}
\usepackage{graphicx}
\usepackage{amsmath,amsfonts,amscd,amssymb}
\usepackage{verbatim}\immediate\write18{texcount -tex -sum  \jobname.tex > \jobname.wordcount.tex}
\usepackage{mathtools} 
\usepackage[customcolors]{hf-tikz} 
\usetikzlibrary{patterns}
\usetikzlibrary{matrix,decorations.pathreplacing}

\pgfkeys{tikz/mymatrixenv/.style={decoration={brace},every left delimiter/.style={xshift=8pt},every right delimiter/.style={xshift=-8pt}}}
\pgfkeys{tikz/mymatrix/.style={matrix of math nodes,nodes in empty cells,left delimiter={[},right delimiter={]},inner sep=1pt,outer sep=1.5pt,column sep=2pt,row sep=2pt,nodes={minimum width=20pt,minimum height=10pt,anchor=center,inner sep=0pt,outer sep=0pt}}}
\pgfkeys{tikz/mymatrixbrace/.style={decorate,thick}}

\newcommand*\mymatrixbraceright[4][m]{
    \draw[mymatrixbrace] (#1.west|-#1-#3-1.south west) -- node[left=2pt] {#4} (#1.west|-#1-#2-1.north west);
}

\newcommand*\mymatrixbracetop[4][m]{
    \draw[mymatrixbrace] (#1.north-|#1-1-#2.north west) -- node[above=2pt] {#4} (#1.north-|#1-1-#3.north east);
}

\author{Maksym V. Zhelyeznyakov}
\affiliation[University of Washington]
{Department of Electrical and Computer Engineering, Seattle, Washington, 98195, USA}
\author{Johannes E. Fr{\"o}ch}
\affiliation[University of Washington]
{Department of Electrical and Computer Engineering, Seattle, Washington, 98195, USA}
\alsoaffiliation[University of Washington]
{Department of Physics, Seattle, Washington 98195, USA}
\author{Anna Wirth-Singh}
\affiliation[University of Washington]
{Department of Physics, Seattle, Washington 98195, USA}
\author{Jaebum Noh}
\affiliation[Pohang University of Science and Technology (POSTECH)]{Department of Mechanical Engineering, Pohang University of Science and Technology (POSTECH), Pohang 37673, Republic of Korea}
\author{Junsuk Rho}
\affiliation[Pohang University of Science and Technology (POSTECH)]{Department of Mechanical Engineering, Pohang University of Science and Technology (POSTECH), Pohang 37673, Republic of Korea}
\alsoaffiliation[Pohang University of Science and Technology (POSTECH)]{Department of Chemical Engineering, Pohang University of Science and Technology (POSTECH), Pohang 37673, Republic of Korea}
\alsoaffiliation[Pohang University of Science and Technology (POSTECH)]
{POSCO-POSTECH-RIST Convergence Research Center for Flat Optics and Metaphotonics, Pohang 37673, Republic of Korea}
\author{Steven L. Brunton}
\affiliation[University of Washington]
{Department of Mechanical Engineering, University of Washington, WA 98195, USA}
\author{Arka Majumdar}
\affiliation[University of Washington]
{Department of Electrical and Computer Engineering, Seattle, Washington, 98195, USA}
\alsoaffiliation[University of Washington]
{Department of Physics, Seattle, Washington 98195, USA}
\email{arka@uw.edu, mzhelyez@gmail.com}

\title
  {Large area optimization of meta-lens\\ via data-free machine learning}

\begin{document}


%
%
%
%
%

\begin{abstract}
Sub-wavelength diffractive optics meta-optics present a multi-scale optical system, where the behavior of constituent sub-wavelength scatterers, or meta-atoms, need to be modelled by full-wave electromagnetic simulations, whereas the whole meta-optical system can be modelled using ray/ wave optics. Current simulation techniques for large-scale meta-optics rely on the local phase approximation (LPA), where the coupling between dissimilar meta-atoms are completely neglected. Here we introduce a physics-informed neural network, which can efficiently model the meta-optics while still incorporating all of the coupling between meta-atoms. Unlike existing deep learning techniques which generally predict the mean transmission and reflection coefficients of meta-atoms, we predict the full electro-magnetic field distribution. We demonstrate the efficacy of our technique by designing 1mm aperture cylindrical meta-lenses exhibiting higher efficiency than the ones designed under LPA. We experimentally validated the maximum intensity improvement (up to $53\%$) of the inverse-designed meta-lens. Our reported method can design large aperture $(\sim 10^4-10^5\lambda)$ meta-optics in a reasonable time (approximately 15 minutes on a graphics processing unit) without relying on any approximation.

\end{abstract}
\vspace{.5em}
\section{Introduction}
In the age of silicon computing, numerical simulations are at the heart of understanding and designing physical systems. For many cases, analytical solutions to complex device geometries are intractable to compute, or simply do not exist. From extremely large systems like rockets \cite{rocket} to ultra-small nanophotonic devices \cite{Molesky2018}, numerical simulations provide scientists and engineers with the necessary tools to design non-intuitive structures. In electromagnetics, direct solvers, including the finite difference time domain (FDTD)\cite{Yee} and the finite difference frequency domain (FDFD) \cite{Rumpf2014,ShinFDFD} simulators, are the usual choices when dealing with heterogeneous structures with sub-wavelength features that require a high degree of numerical accuracy. Most commonly, electromagnetic simulation tools serve to validate the qualitative designs created by engineers based on prior knowledge and intuition. In recent years, the field of nanophotonics has incorporated a new paradigm of computer-aided device design, where a device's performance is summarized by a quantitative figure of merit (FOM) that is optimized over. This method involves running a forward numerical simulation, computing the FOM, and iteratively modifying the device's geometry based on an optimization algorithm to reach the desired FOM. Such optimization methods, often termed as inverse design, have already been used to create novel and efficient nanophotonic structures \cite{Zhan:18,Molesky2018,Piggott2015,Piggott2017,Pestourie:18,Bayati_2020,Zhelyeznyakov21,Elsawy2019,ParkKim,Zhelyeznyakov:20,spectral-spatial-chris}. However, electromagnetic simulators suffer from a computational resource problem when the device dimension becomes large $(\gtrsim 10^3\lambda)$, where $\lambda$ is the device's operating wavelength. As most electromagnetic simulations are performed over a sub-wavelength grid size, with increased size, the number of input variable becomes prohibitively large, making the simulation slow and memory extensive. The limitation of such forward electromagnetic simulators becomes even more severe for inverse design, where many such forward simulations are needed.

Sub-wavelength diffractive optics, also known as meta-optics, present an important testbed for these problems: the constituent elements of the meta-optics, i.e. meta-atoms, are sub-wavelength, but the dimensions of the whole meta-optics are on the order of $\sim 10^3\lambda-10^5\lambda$. Thus the underlying physics of each scatterer has to be modelled using full-wave electromagnetic simulation, but the whole meta-optical system needs to be simulated using ray or wave optics. Such multi-scale electromagnetic simulators invariably rely on approximations, the most common of which is the local phase approximation (LPA): the scattering in any small region is taken to be the same as the scattering from a periodic surface\cite{Pestourie:18}. This approximation allows the simulation of each scatterer in a periodic array, abstracting out the electromagnetic response as a simple phase shift. While this significantly reduces the computational complexity of simulating a meta-optic, this approximation fails to consider the coupling of each scatterer with their dissimilar neighbors. In fact, it has already been shown that meta-optical lenses designed under LPA have sub-optimal efficiencies \cite{Chung:20}, especially when the numerical aperture is large. The LPA becomes even more inaccurate when the material used to create the meta-lens has low index, such as SiN \cite{Bayati:19}. We note that, while a full FDTD coupled with adjoint optimization has been used to design a meta-optic without relying on LPA, their size has been limited to only $\sim 100 \lambda$\cite{Mansouree:20}. LPA can also be bypassed using Mie scattering approaches\cite{zhan:19}, which however limits the shape of scatterers.

To address the computational bottleneck of large-area inverse design, here we introduce a physics-informed neural network (PINN), which can replace a traditional FDTD/ FDFD solver to predict the electric field distribution for a given dielectric distribution. We note that a large number of works already used artifical neural networks to predict spectral responses of meta-optics of varying scatterer geometries~\cite{Malkiel2018,Li:19,Peurifoyeaar4206,kiarashinejad2019,Deeplearningenabledinversedesigninnanophotonics,Liu2018,GaoLi2019,Lin:19,an2019novel}. However, these works used largely periodic structures for which LPA is accurate. We present a solution via PINNs\cite{pinn1,deepxde} for lenses and devices with spatially varying scatterer geometries, where it is necessary to model the whole electric field from several scatterers and their neighbors. PINNs solve partial differential equations (PDEs) by minimizing a loss function constructed from the PDE itself. This loss function is generally some norm of the residual \cite{pinn1} or an energy function derived from the PDE \cite{KARUMURI2020109120}. PINNs have already seen wide usage in the field of fluid mechanics \cite{pinnfluid1,pinnfluid2,pinnfluid3}, biology \cite{Yazdani2020-xa}, and solving stochastic PDEs \cite{ZHANG2019108850}. In electromagnetic inverse problems, PINNs have also been employed to design meta-optics and nanophotonic devices \cite{Chen2020-xz,hpinn}. These works, however, did not demonstrate a simulation speedup, and are limited to the inverse design of very small devices. We also note that pre-trained PINNs have been used to design small gratings\cite{chen2022}; however their methodology is limited to small gratings that deflect light fields to specific angles, and thus cannot be readily used for the inverse design of general meta-optics. 

In our work, we train PINNs to predict the electric fields from a parameterized set of dielectric meta-atoms corresponding to rectangular pillars. We then use this as a surrogate model to design cylindrical meta-lenses operating in the visible with a diameter of 1 mm ($\sim 1500 \lambda$). Large area meta-optics are simulated by partitioning the simulation region into groups of $11$ meta-atoms, with the outermost meta-atoms overlapping.  After simulation, the fields are stitched together. Our PINNs do not require a training data set. They are trained by randomly generating distributions of dielectric meta-atoms $\epsilon$, feeding them into a neural network $NN$, and minimizing the residual of the linear Maxwell PDE operator 
\begin{equation}
||A_{\text{Maxwell}}(\epsilon) NN(\epsilon) - b||_1    
\end{equation}
over the neural network training parameters. This means our PINNs are trained without ever invoking a forward numerical simulation of Maxwell's equations during the training process. Numerical simulations are invoked only to test the neural network performance (see next section and supplement). Once trained, this method can calculate the full electromagntic field response from a 1 mm diameter cylindrical meta-lens at $\sim 630$nm in approximately 3 seconds on a graphics processing unit (GPU). Furthermore, we demonstrate a theoretical and experimental improvement of the maximum intensity of cylindrical metalenses over their forward designed hyperboloid counterparts, signifying the improvement over using LPA. We emphasize that the size of the meta-lens, on which we demonstrate the intensity enhancement of over 50\%, is at least one order of magnitude larger than any other inverse designed lens that does not rely on the LPA. We note that the reported method is robust enough to handle even larger meta-optics, with simulation time scaling only linearly with the aperture of the cylindrical lens (see supplement).
\section{Deep neural network proxy to Maxwell's equations}
\begin{figure}[t!]
\center
\begin{overpic}[width=\textwidth]{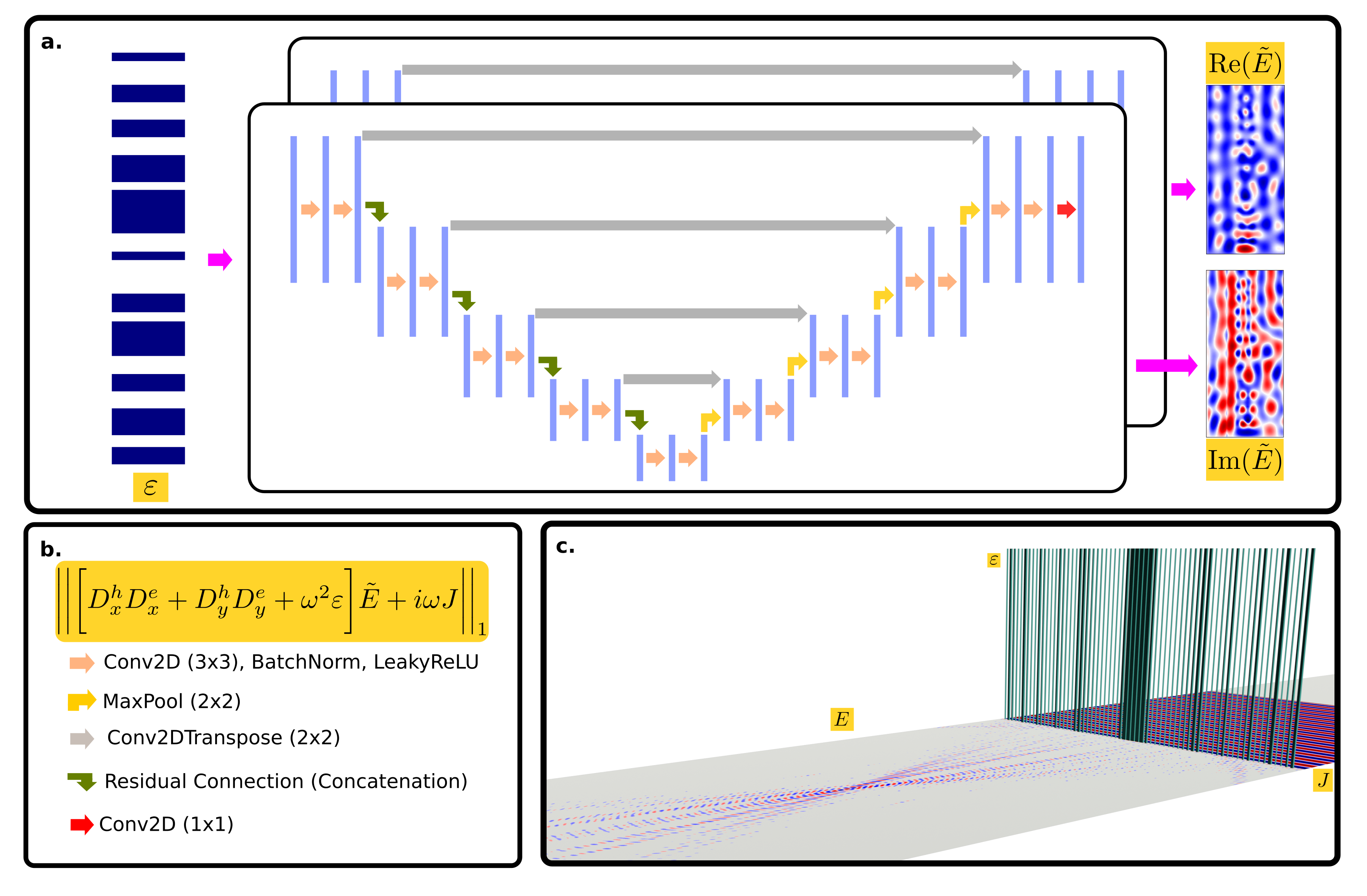}

\end{overpic}
\caption{\textbf{a.} Neural network schematic. $\varepsilon$ distributions of $11$ pillar meta-optics are meshed by randomly generating sets of pillar half-widths of height $h=0.6\mu$m with a dielectric constant $4$ corresponding to SiN. The background medium is air. The loss function is the $||\cdot||_1$ norm of the residual of Eq.~\eqref{eq:3dMaxwell}. \textbf{b.} Neural network architecture. Encoder layers are down-sampled by a maxpool operation with a $2\times2$ kernel. The decoder part of the network is up-sampled by the Conv2DTranspose operation with a $2\times2$ kernel. \textbf{c.} Render of the system under optimization. A current $J$ is incident on a cylindrical metalens with dielectric distribution $\varepsilon$, with output response $E$.}
\label{fig:neuralnetstrategy}
\end{figure}

 Our problem statement is summarized in Fig. \ref{fig:neuralnetstrategy}. The monochromatic electromagnetic scattering equation for an inhomogeneous, non-magnetic material is given by:
\begin{equation}
    \nabla \times \nabla \times \mathcal{E}(x) - \omega^2 \epsilon(x) \mathcal{E}(x) = i \omega \mathcal{J}(x).
\label{eq:3dMaxwell}
\end{equation}
In the 2D case, assuming out of plane polarization $(0,0,\mathcal{E}_z)$, and the double curl vector identity, $\nabla \times \nabla \times = \nabla(\nabla \cdot) - \nabla^2$ we can simplify Eq.~\eqref{eq:3dMaxwell} to:
\begin{equation}
    \nabla^2 \mathcal{E}_z(x) + \omega^2 \epsilon(x) \mathcal{E}_z(x) = -i\omega \mathcal{J}_z
\label{eq:2dMaxwell}
\end{equation}
where $\mathcal{E}_z$ and $\mathcal{J}_z$ are scalar fields. Equation~\eqref{eq:2dMaxwell} is defined over all space, with boundary conditions at $|x| \rightarrow \infty$. To simulate this equation, we discretize it on a Yee grid \cite{Yee} by replacing the $\nabla$ operator with a matrix, and treating the field $\mathcal{E}_z(x)$ and current $\mathcal{J}_z$ as vectors $E$ and $J$ at discrete values of $x$. Similarly, we treat the dielectric distribution $\epsilon(x)$ as a diagonal matrix $\varepsilon$. To truncate the simulation to a finite domain, we use perfectly matched boundary layers (PML), by making the transformation on the partial derivative operators $\frac{\partial}{\partial x} \rightarrow \frac{1}{1+i\frac{\sigma(x)}{\omega}}\frac{\partial}{\partial x}$. Making these substitutions, Eq.~\eqref{eq:2dMaxwell} becomes:
\begin{equation}
    \bigg[D^h_xD^e_x+D^h_yD^e_y+\omega^2 \varepsilon \bigg] E = -i\omega J
\label{eq:matrixMaxwell}
\end{equation}
with matrices $D^h_x, D^e_x, D^h_y, D^e_y$ being the matrix representations of corresponding derivative operators on a Yee grid with incorporated PML boundaries. See supplement \ref{s-fdfd} for a more detailed description of the matrices. These matrices were extracted from a modified version of the package angler \cite{hughes2018adjoint} with constants $c,\epsilon,\mu$ set to 1 and the length scale set to $\mu m$. To build a neural network proxy to solve Eq.~\eqref{eq:matrixMaxwell}, we employ a PINN. PINNs generally use the coordinates of the computational grid as the input to the neural network, and then minimize the residual of the physical equations by approximating the target quantity being solved for with a neural network. This approach is slow since it effectively functions as an iterative solver re-parametrized over neural network weights and biases. It also required retraining the neural network for all different dielectric distributions. Our approach is to build a proxy solver that predicts the field $E$ from a dielectric distribution $\varepsilon$. We pre-train the PINN to predict fields from inputs $\varepsilon$ before optimizing our meta-lenses. The minimization problem to train the PINN becomes:
\begin{equation}
\min_\theta f(\varepsilon; \theta)\quad \text{where} \quad
    f(\varepsilon; \theta) = \bigg|\bigg| \big[ D^h_xD^e_x+D^h_yD^e_y+\omega^2 \varepsilon \big] NN(\varepsilon; \theta) +i\omega J\bigg|\bigg|_1
\label{eq:pinnloss}
\end{equation}

\begin{figure}[t]
\begin{overpic}[width=\textwidth]{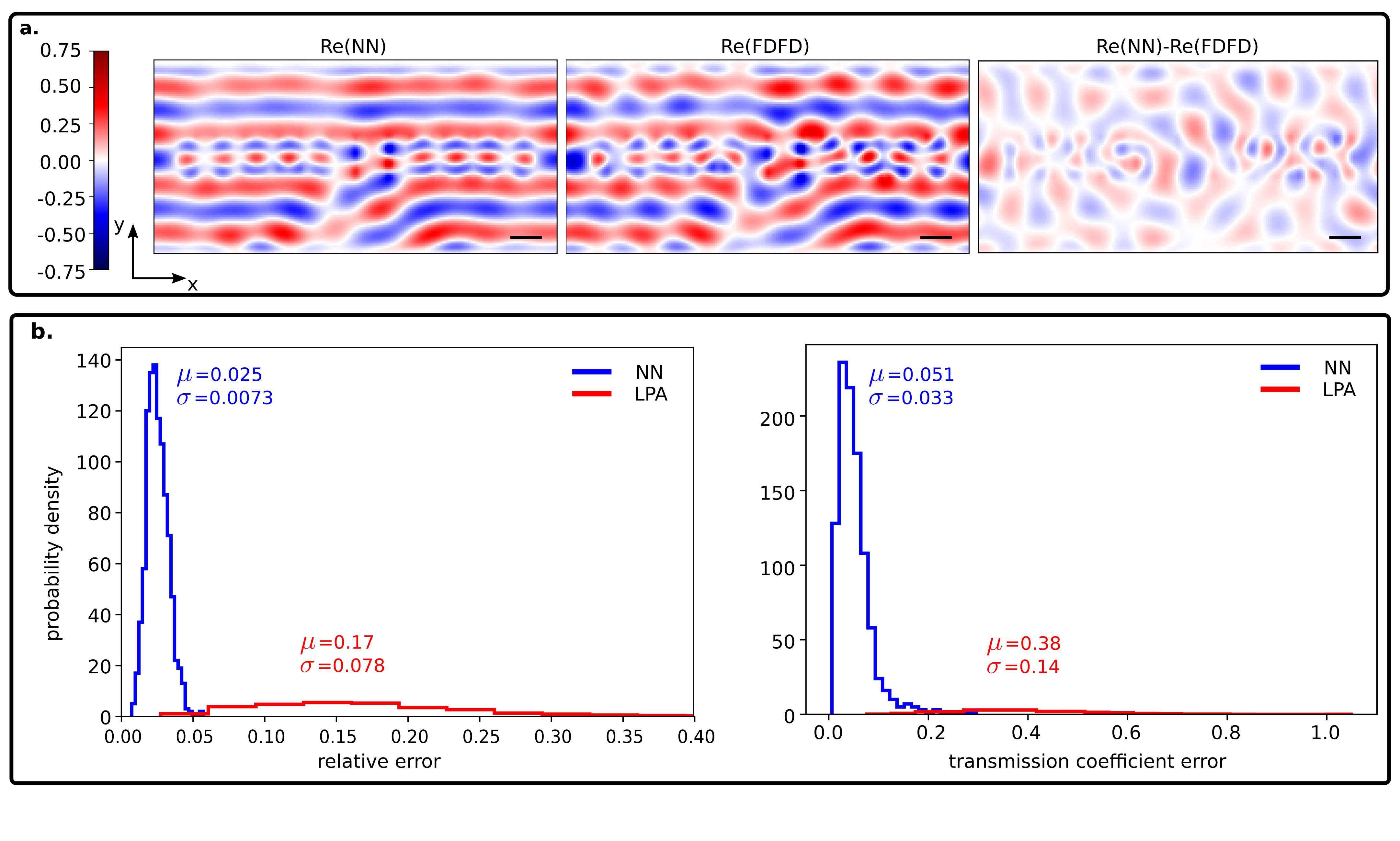}
\end{overpic}
\caption{\textbf{a.} Real part of fields predicted by (left) neural network, (center) FDFD, and (right) the difference between the true and predicted fields. \textbf{b.} Comparison between the performance of the proposed neural network and LPA methods. (left) Shows the relative error between FDFD predicted fields and the fields predicted by LPA. (right) Error comparisons between the transmission coefficients predicted by LPA and the neural net.}
\label{fig:exampleErrors}
\end{figure}

\noindent
with $NN(\varepsilon; \theta)$ being the output field from the PINN, and $|| \cdot ||_1$ is the vector $l_1$ norm. Here $\theta$ refers to the weights and biases of the neural network $NN$. A lower physics informed loss indicates that the neural network is actually satisfying the PDE, and thus predicting the field more accurately. We re-emphasize that there is no data term in $f(\epsilon;\theta)$, which simplifies the neural network training process. Furthermore, we believe that it mitigates the accumulation of error in the gradients during the inverse design process observed by Chen et. al. \cite{chen2022}. Fig. \ref{fig:neuralnetstrategy} outlines the general strategy for building the proxy model. During each epoch, 10 (batch size) dielectric distributions consisting of rectangular pillars of height $h=0.6 \mu m$ with dielectric constant $4$ (corresponding to SiN), are generated from $11$ random pillar half-widths per batch. The operation wavelength is $\lambda=0.633 \mu$m. The neural network architecture chosen is a UNET, shown in Fig. \ref{fig:neuralnetstrategy} \textbf{a} and \ref{fig:neuralnetstrategy} \textbf{b}, due to its relatively good performance with scattering problems\cite{chen2022}. The model is trained for $5\times 10^5$ epochs using the ADAM optimizer \cite{adam} with a learning rate set to $5\times 10^{-4}$. The final residual of the fields predicted by the neural network are of the order of $\sim 0.5$, compared to the numerical residual produced by FDFD which is on the order of $10^{-16}$. Although there is a large difference, in the next section we show that this still produces a simulator which is capable of outperforming the LPA when optimizing the efficiency of a metalens. Fig. \ref{fig:exampleErrors} \textbf{a} shows an example of a field predicted from a random set of pillars by the neural network, by a 2D FDFD code, and their difference, showing good qualitative match. A more quantitative measure at the errors is shown in Fig. \ref{fig:exampleErrors} \textbf{b}, where we show the point-wise error probability density functions for the relative error between the complex fields predicted by FDFD and that predicted by neural network and field predicted under LPA, and the absolute error between pillar-wise average transmission coefficients. See supplement \ref{s-pillar_wise_transmission_coeff} for a more detailed description of the pillar wise transmission coefficient error. The relative error is expressed as: 
\begin{equation}
    \text{mean}\bigg(\frac{|E_{\text{approx}}-E_{\text{FDFD}}|^2}{\max(|E_{\text{FDFD}}|^2)}\bigg).
\label{eq:relative_error}
\end{equation}
For the PINN, $E_{\text{approx}}$ is the field predicted from a set of 11 pillars. For the LPA, $E_{\text{approx}}$ is fields predicted from the same set of pillars, and then stitched together over the same region. See fig. \ref{fig:tcoeffs} for a visual explanation. This error can be interpreted as the $\%$ difference between the FDFD predicted field and the approximate field, either predicted by neural network without making any LPA or the field under LPA. The mean expected relative error for the neural network is $\mu = 0.025$ with a standard deviation of $\sigma = 0.0073$. When using the LPA over the same region, we get a mean relative error of $\mu=0.17$ with a standard deviation of $0.078$. Thus, based on the relative field error, our method is 6.8$\times$ more accurate than the LPA. For the pillar-wise transmission coefficient error, we get an expected error of $\mu = 0.051$ for the neural network with a standard deviation of $\sigma = 0.033$ and for the LPA method we get an expected error of $\mu = 0.38$ with a standard deviation of 0.14. Thus, based on the transmission coefficient error, our method is 7.2$\times$ more accurate than the LPA. 
\section{Device optimization}
\begin{figure}[t]
\begin{overpic}[width=\textwidth]{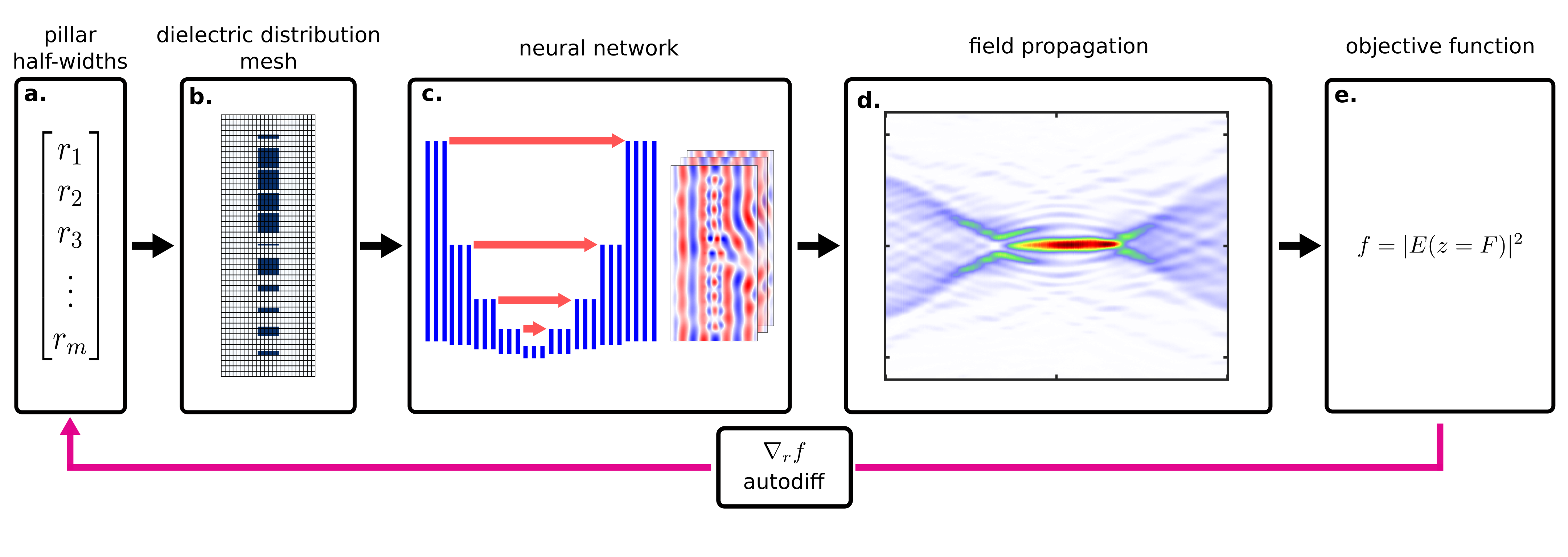}
\end{overpic}
\caption{Optimization strategy of 2D meta-optics with PINNs. \textbf{a.} We start with a vector, which contains a list of all pillar half-widths, characterizing the meta-optic. These half-widths are then batched into groups of 11 with an overlap of 1 pillar on each side. The choice of $11$ pillars was made based on the GPU memory required to train the PINN. \textbf{b.} The half-widths are meshed into dielectric distributions which get fed into the \textbf{c.} The neural network predicts patches of fields which are then stitched together, and \textbf{d.} propagated via the angular spectrum method. \textbf{e.} The objective function is formed from the resulting field, and backpropagated using PyTorch's automatic differentiation functionality to update the initial radius distribution.}
\label{fig:optimization}
\end{figure}
The optimization process based on automatic differentiation functionality of PyTorch for large area meta-optics is outlined in Fig. \ref{fig:optimization}. The forward problem is solved via a pre-trained PINN. Since the input into the neural net is a meshed grid of pillars, a differentiable map from pillar half-widths to meshed geometries must be generated. This is achieved by generating Gaussian functions centered around pillar centers, with standard deviations of pillar half-widths in the x dimension, and pillar height in the y dimension, and then using a modified softmax function to transform the Gaussians into rectangles with slightly rounded edges, making them differentiable via automatic differentiation (see supplement \ref{problem-setup}). The meshed structures are fed into two separate neural networks that have been pre-trained to predict the complex electric field. The fields are then stitched together with regions of the outer half-widths overlapping. The total field is then propagated using the angular spectrum method. The propagated field is used to calculate the FOM $f$ from Eq.~\eqref{eq:pinnloss}. We use automatic differentiation to compute the gradients of the FOM with respect to the input half-widths $\nabla_{\vec{r}} f$, and iteratively update them with the ADAM optimizer\cite{adam}.
\section{Results}
We used PINN surrogate model to optimize $9$ different lenses, all with $1$ mm aperture, with focal lengths ranging from 250-1500 $\mu$m in increments of 250 $\mu$m. The minimum feature size is set to 75 $nm$, to ensure fabricability. To compare our optimization approach, we also generated lenses according to the hyperboloid phase equation:
\begin{equation}
    \phi(x,y) = \frac{2\pi}{\lambda}\big(\sqrt{x^2+F^2}-F\big)
\end{equation}
\begin{figure}[t]
\begin{overpic}[width=\textwidth]{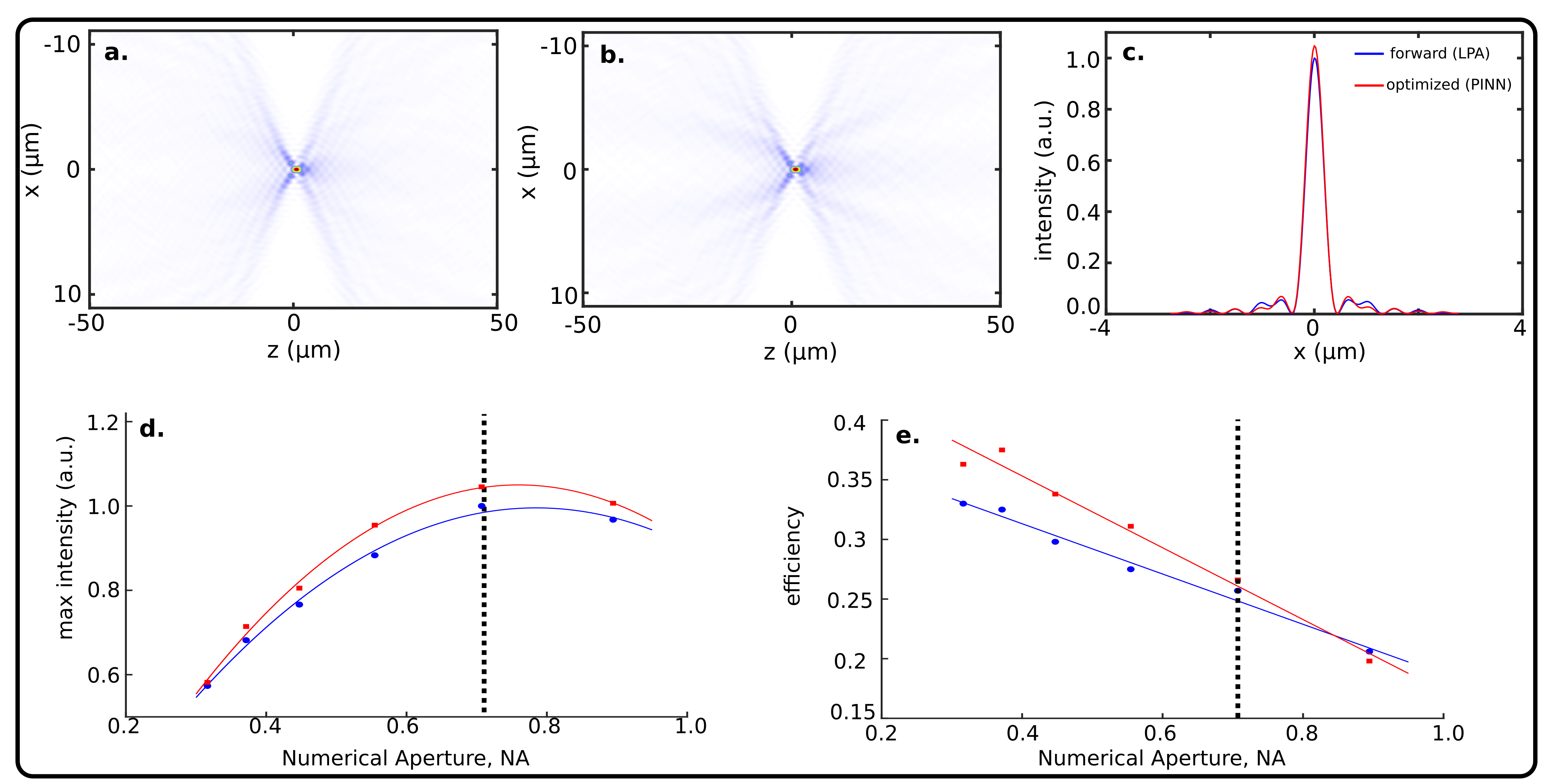}

\end{overpic}
\caption{Efficiency and intensity sweeps of forward designed lenses and optimized lenses. \textbf{a.} Focal spot intensity profile of a forward designed lens with focal length $F=500 \mu$m. \textbf{b.} Focal spot intensity profile of an optimized lens. \textbf{c.} Slices of intensity profiles for both lenses. The intensity was normalized such that the maximum intensity of the forward designed lens is 1. The theoretical performance improvement is $\sim$ 3\%. \textbf{d.} Maximum intensity at the focal spot vs lens numerical aperture (NA). Intensities are normalized such that the maximum of the largest forward designed intensity is set to 1. \textbf{e.} Theoretically computed efficiencies of the lenses vs NA. The solid lines are visual aids for the trend and do not correspond to a theoretical prediction.}
\label{fig:eff_imax}
\end{figure}
The phase is implemented under LPA using SiN (refractive index 2), a wavelength of $0.633\mu$m, and periodicity of $p=0.443\mu$m. We then optimize the lens employing our PINN to increase the intensity at the focal spot, i.e., the FOM is given by:
\begin{equation}
    f = \big|E(x=0,z=F)\big|^2
\end{equation}
Fig. \ref{fig:eff_imax}\textbf{a} and Fig. \ref{fig:eff_imax}\textbf{b} show the intensity profile of a forward designed and optimized lens with $F=500\mu$m focal length.  Fig.\ref{fig:eff_imax}\textbf{c} shows the normalized intensity slice at the focal spot of both lenses.
As seen in Fig. \ref{fig:eff_imax} \textbf{d} the maximum intensities at the focal spots improve in every case. Fig. \ref{fig:eff_imax} \textbf{e} shows that the efficiency all improves in all except for the lens with highest NA.
\begin{figure}[t]
\begin{overpic}[width=\textwidth]{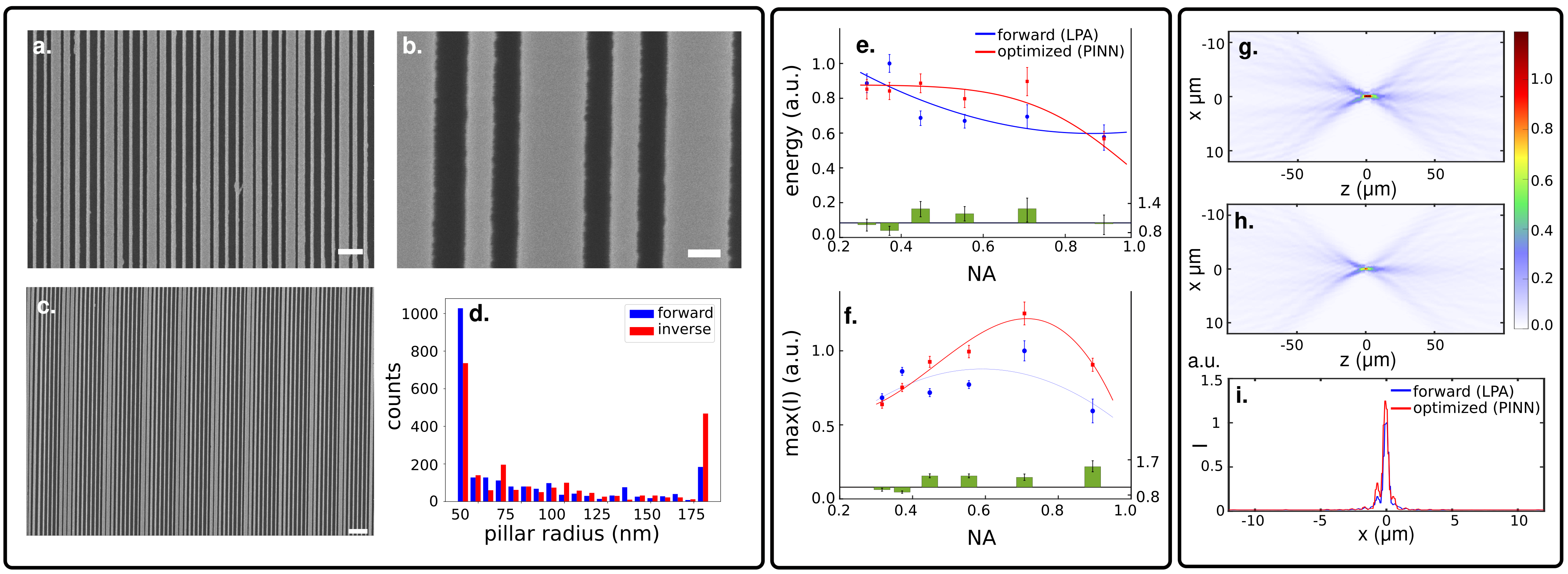}

\end{overpic}
\caption{\textbf{a.-c.} Scanning Electron Microscope (SEM) images of the fabricated SiN meta-lens with focal length $F=500\mu$m. The scale bars correspond to $1\mu m$, $0.1\mu m$, and $1\mu m$, respectively. \textbf{d.} Counts of pillar half-widths of the forward and inverse designed lens. \textbf{e.} Measured intensity contained in the region given by $3\times$ FWHM of the focal spot vs lens numerical aperture. The units are normalized to the largest intensity integral of the forward design. \textbf{f.} Maximum intensity at the focal spot. The inverse designed lenses outperform the forward designed lenses for NA > 0.44. The lines are visual aids and not fits to a theoretical model. Units are normalized to the largest intensity of the forward designed lens. In the NA $= 0.9$ ($F=250\mu$m) case an improvement of $53\%$ is observed. \textbf{g.} Experimentally measured field intensities of the forward designed lens and \textbf{h.} of the inverse designed lens. \textbf{i.} Intensity slice at the focal spot. The intensities are normalized such that the maximum intensity of the forward designed lens is 1. The intensity of the inverse designed lens focal spot is 1.25.}
\label{fig:experiment}
\end{figure}

We validated our designs by fabricating and experimentally testing the meta-lenses using a microscope (details of fabrication and characterization in Methods). Fig. \ref{fig:experiment} shows an example of the inverse optimized device. Fig. \ref{fig:experiment}\textbf{a-c} shows the scanning electron micrographs (SEMs) of the fabricated optimized lens with focal length $F=500\mu$m. Fig.\textbf{5d} shows the distribution of the dielectric pillar half-widths of the same forward and optimized lens. signifying the two designs are very different. Fig.\ref{fig:experiment}\textbf{e} shows the focal spot intensities of the lenses integrated over a $r=3\times$FWHM region at the focal spot, which yields a quantitative value to compare the lens efficiency \cite{faraonefficiency} among different devices. Fig.\ref{fig:experiment}\textbf{f} plots the maximum intensity plot as a function of the lens NA. For optimized lenses with NA > 0.44, we see improvements of more than $25\%$, with a maximum improvement of $53\%$ for the NA = 0.9 lens. The experimentally determined intensity integral, which is analogous to the efficiency of a lens, on has improvements of more than $18\%$ in all cases except for the NA=0.9 case. This is because the FWHM of the optimized lens at the NA=0.9 case is actually smaller than the FWHM of the forward designed lens, leading to a smaller integration area when computing the energy.

 Fig. \ref{fig:experiment}\textbf{g} shows experimentally measured field profiles of the forward designed $F=500\mu$m meta-lens. Fig. \ref{fig:experiment}\textbf{h} shows the same for an optimized lens. Fig.\ref{fig:experiment}\textbf{i} is the slice of the focal spot intensity profile along the $z=F$ plane. In all these figures, the intensity is normalized such that the maximum intensity of the forward designed lens is 1.
\section{Discussion}
We have developed a PINN to use as a proxy surrogate model for simulating the full Maxwell's equations to design dielectric meta-optics. We used the PINN to optimize pillar half-widths to maximize the intensity at the focal spot of 1 mm aperture cylindrical meta-lenses. We demonstrated experimental improvements of the maximum intensity of the lenses up to $53\%$. We also want to note that this method was useful for the inverse design of extended depth of focus lenses\cite{Bayati_2020} (see supplement \ref{s-edof}). This model did not use the LPA, but simulated meta-atoms by splitting up the device into chunks with overlapping boundaries, and stitching the chunks together to approximate the full field response.  We emphasize that FDFD simulations were never carried out to train the PINN, and we only minimized the residual of the PDE itself to train the network. The PINN training took approximately 2 hours on our machine. In our studies, this method provided approximately a 3-5x speedup over conventional FDFD with overlapping boundary conditions, and was much simpler to use as a forward simulator for optimization problems since it can be used as a simple map from $\epsilon$ to $E$-field with gradients computed by automatic differentiation. 

We would also like to note that the theoretical intensity and efficiency improvements are quite a bit smaller than their experimental counterparts. While we don't have a clear explanation for this discrepancy, the theoretical and experimental trends in lens improvement are similar. We hypothesise that the inverse designed lenses may be more tolerant to fabrication imperfections. The inverse design solution we introduced in this paper can be integrated into various computationally intensive tasks which require mate-optical inverse design such as the end-to-end optimization of computational imaging systems and the design of optical neural networks\cite{Tseng2021NeuralNanoOptics,d2nn}. It is worth noting, however, that this method is not a general numerical solver. It is limited to predicting electromagnetic field responses from fixed source, material, and boundary parameters. Source type and $k$-vector, dielectric constant, geometry type (rectangular pillar of fixed height), and boundary conditions must all remain constant for this method to work. If any of these parameters are modified, the PINN must be retrained. Furthermore, the method we presented was only implemented under a 2D approximation. Extending this method to 3 dimensions would take significant effort due to the fact that the electric field $\mathcal{E}$ could no longer be treated as a scalar field, and the full vector nature would have to be modeled. On a $n \times n$ grid in 2D, the Maxwell operator $\big[\nabla^2+\omega^2 \epsilon\big]$ results in a $n^2 \times n^2$ matrix, while for a $n\times n \times n$ 3D grid the Maxwell operator $\big[\nabla\times\nabla\times - \omega^2 \epsilon]$ result in $9n^3 \times 9n^3$ square matrices due to the additional 2 vector field components that must be modeled. However, these operators are sparse with number of nonzero elements that scale as $\sim 38n^3$ in 3D, making small problems still manageable. The other problem with generalizing this method to 3D is the large null-space of the $\nabla\times\nabla\times$ operator which results in slow convergence of numerical methods\cite{curl-curl-1,curl-curl-2}. Its highly likely that this could also affect the training of the PINN, and require regularization or preconditioning which deflates the null space of this operator to properly converge onto a solution. On the other hand, in this work we showed that machine-precision numerical accuracy of numerical solvers may be not be needed for inverse design methods with FDFD. Solvers could be sped up by relaxing the relative error tolerance, such that iterative solvers converge quicker for predicting the forward and adjoint problems. In future work we aim to explore these options.

\section{Methods}
\subsection{Fabrication}
All devices described and discussed in Figure 5 (forward and PINN designed) were fabricated on the same substrate. First a $\sim$ 700 nm SiN film was deposited on a quartz wafer using plasma enhanced chemical vapor deposition (SPTS Delta LPX PECVD). A thin film of a polymer resist (ZEP 520-A) and a thin film of a discharging polymer layer (DisCharge H2O) were subsequently spun onto the sample. We then used electron beam lithography (JEOL JBX-6300FS, 100 keV, 2nA) to write the various structures. After development, a short descum step (Glow Research, Autoglow, 12 s, 100 W) was used to remove remaining resist residues and subsequently a layer of ~ 60 nm AlOx was deposited using a home-built e-beam evaporator. After overnight lift-off in warm NMP and a further plasma cleaning step in O2, we used inductively coupled reactive ion etching (Oxford Instruments PlasmaLabSystem100) with a fluorine gas chemistry to transfer the pattern from the AlOx hard mask into the underlying SiN layer. The final thickness of the etched layer indicated a pillar height of $\sim$650 nm. 

\subsection{Experiment set up}
For intensity measurements, light from a HeNe laser was transmitted through the backside of the chip and measured on the device side using a translatable microscope relay setup. In detail, the sample was mounted at a fixed position on a kinematic holder, allowing the fine adjustment for pitch and yaw, as well as the lateral position. Light was transmitted through the substrate side and would propagate entirely in air. The resulting focusing pattern was measured using a microscope setup consisting of a Nikon 100X LU Plan Fluor objective with 0.9 NA (equal or higher to the NA of the meta-optic), a tube lens (Thorlabs), and a camera (Allied Vision ProSilica GT1930), which were mounted on a programmable automated translation stage (NewPort). Frames were acquired at specific intervals of the movement, which allowed the reconstruction of the intensity vs focal distance.      
\subsection{Machine specifications}
Ubuntu 22.04
\newline
\noindent
2x Intel E5-2620 at 2.1 GHz
\newline
\noindent
NVIDIA Tesla K40 12 GB Memory running CUDA 11.4
\newline
\noindent
128 GB DDR3 Memory
\section{Acknowledgements} 
This research was funded by NSF-GCR-2120774. M.V.Z. is supported by an NSF graduate fellowship. Part of this work was conducted at the Washington Nanofabrication Facility/Molecular Analysis Facility, a National Nanotechnology Coordinated Infrastructure (NNCI) site at the University of Washington, with partial support from the National Science Foundation via Awards NNCI-1542101 and NNCI-2025489.
\section{Associated Content}
Code for the project is available at \text{https://github.com/demroz/pinn-ms}
\bibliography{mybib}

\pagebreak
\begin{center}
\textbf{\large Supplemental Materials: Large area optimization of meta-lens\\ via data-free machine learning}
\end{center}
\newpage
\setcounter{equation}{0}
\setcounter{figure}{0}
\setcounter{table}{0}
\setcounter{section}{0}
\makeatletter
\renewcommand{\theequation}{S\arabic{equation}}
\renewcommand{\thefigure}{S\arabic{figure}}
\renewcommand{\bibnumfmt}[1]{[S#1]}
\renewcommand{\citenumfont}[1]{S#1}
\renewcommand{\thesection}{S\arabic{section}} 

\section{Problem Setup}
\label{problem-setup}
\begin{figure}[t]
\center
\begin{overpic}[width=\textwidth]{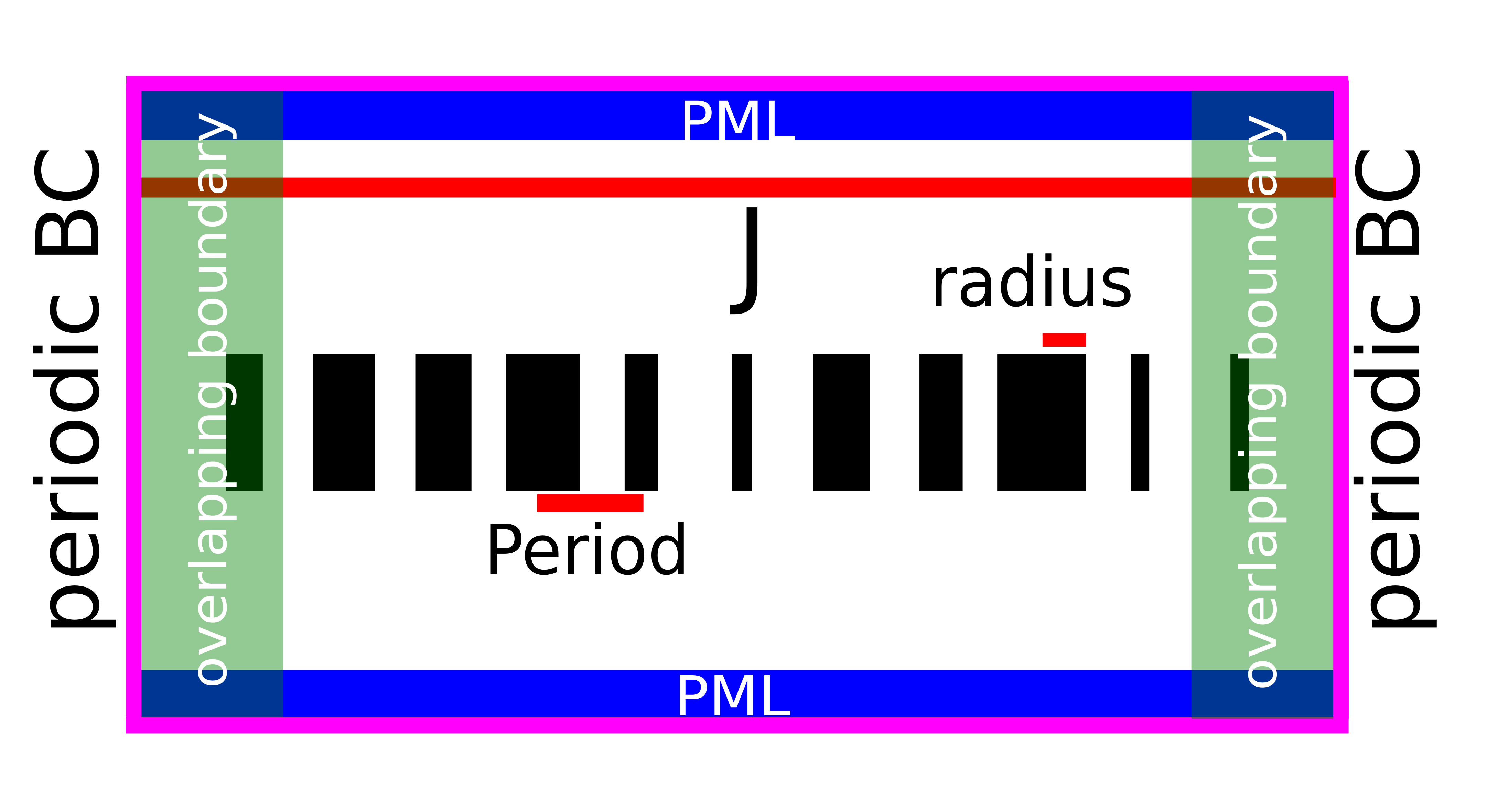}
\end{overpic}
\caption{Simulation problem setup}

\label{fig:problem setup}
\end{figure}

The neural networks are trained to predict electric field responses of distributions of dielectric scattereres, here SiN pillars, from a plane wave current source of wavelength $\lambda=633nm$. The resolution is set to be 16 pixels per period, with each period being  $0.443\mu$m. The boundary conditions along the x-direction are set to be periodic. The boundary conditions along the y-direction are set to be 10 grid points of a perfectly matched layer (PML). The simulation domain in the x direction is $12\times$period = $5.316\mu$m, and $6\times$ period = $2.658\mu$m along the y direction. When simulating field responses of large area metasurfaces, the simulation region is split up into groups of $11$ periods with the outermost periods overlapping. Each region is simulated with a trained neural network as shown in \ref{fig:problem setup}. The field responses of the innermost 9 pillars are stitched together. 
\newpage
\section{Pillar-wise transmission coefficient error}
\label{s-pillar_wise_transmission_coeff}
\begin{figure}[t]
\center
\begin{overpic}[width=\textwidth]{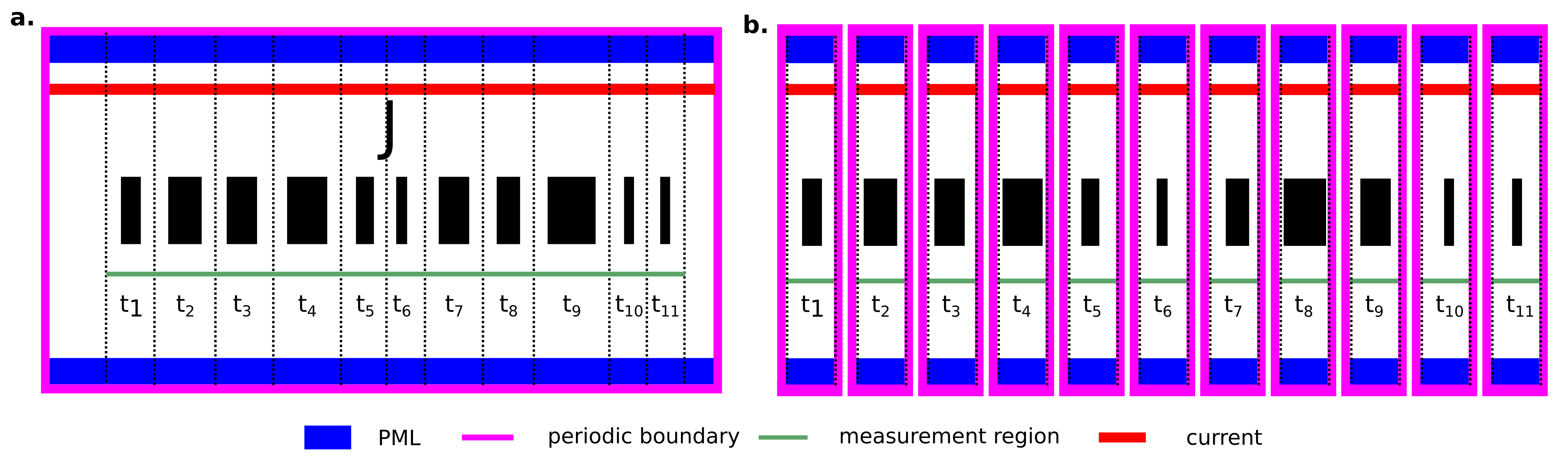}
\end{overpic}
\caption{Calculating pillar-wise transmission coefficients for \textbf{a.} FDFD and neural network simulations and \textbf{b.} LPA simulations.}
\label{fig:tcoeffs}
\end{figure}
Here, we summarize how we compute the average pillar-wise transmission error (i.e. main text Figure 2 \textbf{b} right hand side). For every set of 11 pillars we compute the transmission coefficients for each pillar using FDFD, PINNs, and LPA. Fig. \ref{fig:tcoeffs} (left) shows how transmission coefficients for each batch are computed for FDFD and PINN. Fig. \ref{fig:tcoeffs} (right) shows how transmission coefficients are computed under the LPA. For the FDFD and PINN case, we simulate the full field over the simulation region defined in the previous section. Then we measure field 6 pixels (0.13$\mu$m) away from the meta-atom. Then that field is averaged over a single period corresponding to the location of the meta-atom. This gives us 11 transmission coefficients. For the LPA, we simply simulate the 11 scatterers under periodic boundary conditions, and extract the average field the same distance away from each pillar. The mean transmission coefficient error is thus given by
\begin{equation}
    \frac{1}{11}\sum_{i=1}^{11} |t_i^{FDFD}-t_i^{\text{approx}}|^2
\end{equation}
where $t_i^\text{FDFD}$ is the transmission coefficient computed with FDFD and $t_i^\text{approx}$ is computed by either the neural network or LPA.

\newpage
\section{Mesh reparametrization}
\label{s-mesh-reparametrization}
\begin{figure}[t]
\center
\begin{overpic}[width=\textwidth]{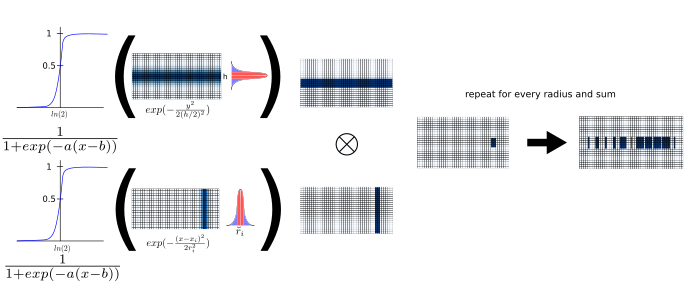}
\end{overpic}
\caption{transforms carried out to make a differentiable map from $r \rightarrow \epsilon$}

\label{fig:transforms}
\end{figure}
To mesh a set of pillars, we first generate a grid of coordinates $x \in [-6p, 6 p]$  and $y \in [-\frac{3}{2}p, \frac{3}{2}p ]$, $p$ being the pitch of the meta-optics. Since we have a fixed set of pillars, their centroid locations are given by $x_i = \frac{n p }{2}$ where $n$ is the pillar index running from $-5$ to $5$. The first transform we define is just a shifted gaussian function on this grid:
\begin{equation}
    T_1 (r_i, x_i) = \text{ex}p\big(-\frac{(x-x_i)^2}{2 r_i^2}\big)
\end{equation}
Similarly, the second one defines the height:
\begin{equation}
    T_2 = \text{exp}\big(-\frac{y^2}{2 (h/2)^2}\big)
\end{equation}
The modified softmax is defined by:
\begin{equation}
    T_3(a,b,x) = \frac{1}{1+\text{exp}(-a(x-b))}
\end{equation}
$a$ denotes the aggressiveness of the softmax function, and $b$ is the point above which the function goes to 1, and below which the function goes to 0. Here, we chose $a=100$ and $b=\text{log}_e(2)$. The aggressiveness was experimentally determined, with larger values causing gradients to become too steep, and lower values causing things to not resemble pillars as much. $b=\text{log}_e(2)$ is chosen because a gaussian function drops to $\text{log}_e(2)$ of its max values after 1 standard deviation of its input variable, hence creating pillars of size $r$. Thus, 1 meshed batch of radii can be written as:
\begin{equation}
    \sum_{i=1}^{11} T_3(T_1(r_i,x_i),100,\text{log}_e(2)) T_3(T_2,100,\text{log}_e(2))
\end{equation}
\newpage
\section{Form of FDFD linear system}
\label{s-fdfd}
\begin{figure}[t]
\center
\begin{overpic}[width=\textwidth]{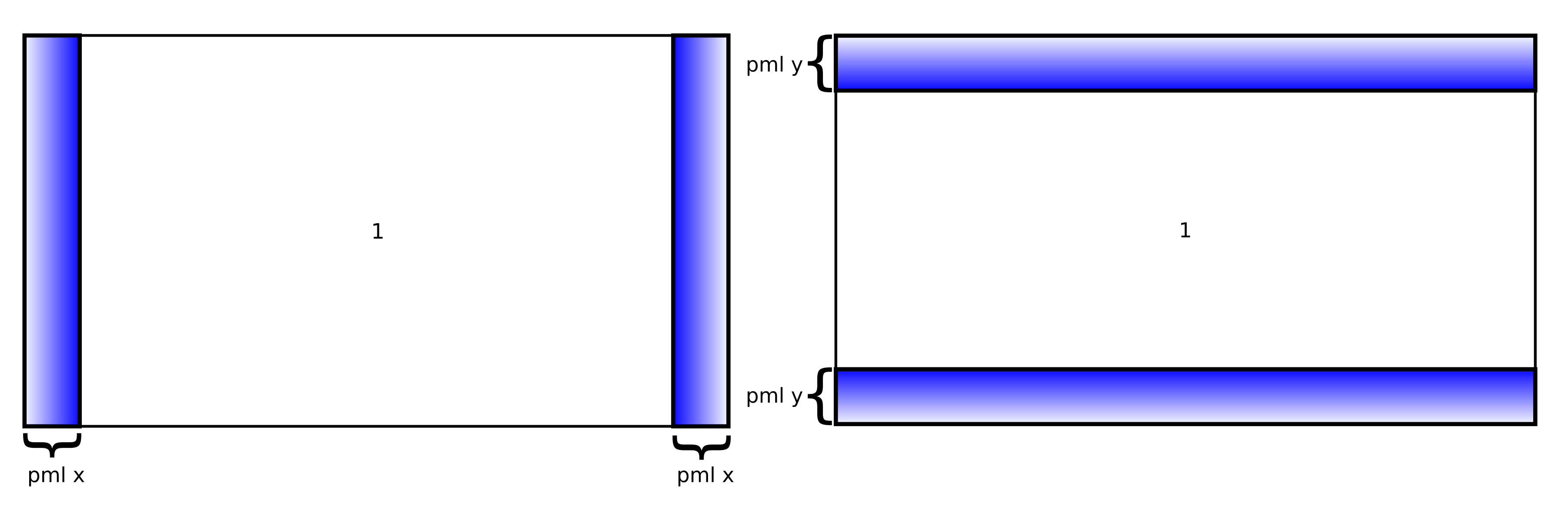}
\end{overpic}
\caption{Description of setup for scale matrices $S_x$ and $S_y$. Left hand side is for the x derivative and right hand side is for the y derivative. PML scaling is computing on a meshed grid, then flattened and embedded in a diagonal matrix.}
\label{fig:pml}
\end{figure}
In this section we give a brief summary of formulating the FDFD linear system on a Yee grid for completeness. The full details of the method can be found elsewhere \cite{sref1, sref2, sref3, sref4, sref5}, so we only give a brief description here. The boundary conditions of Maxwell's equations are defined at $|x|\rightarrow \infty$, so perfectly matched layer (PML) boundary conditions must be implemented to truncate the simulation region to a finite size. To do this, scale matrices $S_x$ and $S_y$ need to be generated. They are constructed by creating scale factors
\begin{equation}
s_w(l) = \begin{cases}
1-i\frac{\sigma_w(l)}{\omega_o \epsilon_0} &\text{inside $w$ normal pml}\\
1 &\text{otherwise}
\end{cases}
\end{equation}
Where $w$ is the coordinate normal ($x$ or $y$), $l$ is the distance inside the PML from the PML interface, $\omega_o$ is the operating angular frequency, and $\epsilon_0$ is the permittivity of free space. $\sigma_w(l)$ is given by
\begin{equation}
    \sigma_w(l) = \sigma_{w,\text{max}} \big(\frac{l}{d}\big)^m
\end{equation}
and $\sigma_{w,max}$ is
\begin{equation}
    \sigma_{w,\text{max}}  = \frac{(m+1) \text{lnR}}{2\eta_0 d}
\end{equation}
$d$ is the thickness of the PML. $\eta_0$ is the vacuum impedance $\eta_0 = \sqrt{\mu_0/\epsilon_0}$. R is the target reflection coefficient. A good reference for PML boundaries is Shin et. al.\cite{sref4}. The package we used, angler\cite{sref3}, uses the convention $m=4$ and ln(R) = $-12$. Furthermore we modified the constants inside the package such that $\mu_0=\epsilon_0=\eta_0=1$. The matrices $S_x$ and $S_y$ are created by computing $s_{w}$ on a meshed grid, flattening it, and embedding it into a diagonal matrix:
\begin{equation}
    S_w =\begin{bmatrix}
    \frac{1}{s_w^{1,1}} & 0 & 0 & \cdots & 0 \\
    0 & \frac{1}{s_w^{2,2}} & 0 & \cdots & 0 \\
    0 & 0 & \frac{1}{s_w^{3,3}} & \cdots & 0 \\
    \vdots & & & \ddots & & \\
    0 & 0 & 0 & \cdots & \frac{1}{s_w^{N,N}}
    \end{bmatrix}
\end{equation}
The numerical derivative matrices on a yee grid, with periodic boundary conditions $\Delta_x$ and $\Delta_y$ are:
\newline

\begin{equation}
    \mathbf{\Delta_x} = 
    \begin{tikzpicture}[baseline={-0.5ex},mymatrixenv]
        \matrix [mymatrix,inner sep=4pt] (m)  
        {
        -\frac{1}{dx} &  & & \frac{1}{dx} &  \\
  & -\frac{1}{dx} & &  & \frac{1}{dx} \\
 &  & \ddots &  &  & \ddots \\
 &  &  & -\frac{1}{dx} &  &  & \frac{1}{dx} \\
 &  &  &  & -\frac{1}{dx} &  &  & \frac{1}{dx} \\
\frac{1}{dx} &  &  &  &  & -\frac{1}{dx} &  &  \\
 & \ddots &  &  &  &  & \ddots \\
 &  & \frac{1}{dx} &  &  &  &  & -\frac{1}{dx}   
 \\
        };

        \mymatrixbraceright{1}{6}{$(N_x-1)N_y$}
        \mymatrixbracetop{1}{4}{$N_y$}
    \end{tikzpicture}
\end{equation}

\begin{equation}
    \mathbf{\Delta_y} = 
    \begin{tikzpicture}[baseline={-0.5ex},mymatrixenv]
        \matrix [mymatrix,inner sep=4pt] (m)  
        {
        -\frac{1}{dy} & \frac{1}{dy} & & & & \\
  & -\frac{1}{dy} & \frac{1}{dy} &  \\
 &  & \ddots & \ddots \\
\frac{1}{dy} &   &  & -\frac{1}{dy} & 0 \\
 &   &  &  & -\frac{1}{dy} & \frac{1}{dy} &  \\
 &  &  &  &   & -\frac{1}{dy} & \frac{1}{dy} &  \\
 &  &  &  &  &  &  & \ddots & \ddots \\
  & &  &  &  &  & \frac{1}{dy} &  &  & -\frac{1}{dy}  
 \\
        };

        \mymatrixbraceright{1}{4}{$N_y$}
        \mymatrixbracetop{1}{4}{$N_y$}
    \end{tikzpicture}
\end{equation}
here $N_x$ and $N_y$ are the grid sizes in the $x$ and $y$ directions and $dx$ and $dy$ are the grid spacings. The derivative matrices $D_x^e$, $D_y^e$, $D_x^h$, and $D_y^h$ can then be constructed as:
\begin{equation}
    D_x^e = S_x \Delta_x
\end{equation}
\begin{equation}
    D_y^e = S_y \Delta_y
\end{equation}
\begin{equation}
    D_x^h = S_x (-\Delta_x^\dagger)
\end{equation}
\begin{equation}
    D_y^h = S_y (-\Delta_y^\dagger)
\end{equation}
where the $^\dagger$ operator is the Hermitian transpose.
\newpage
\section{Neural network training}
\label{s-nn-training}
\begin{figure}[t]
\center
\begin{overpic}[width=\textwidth]{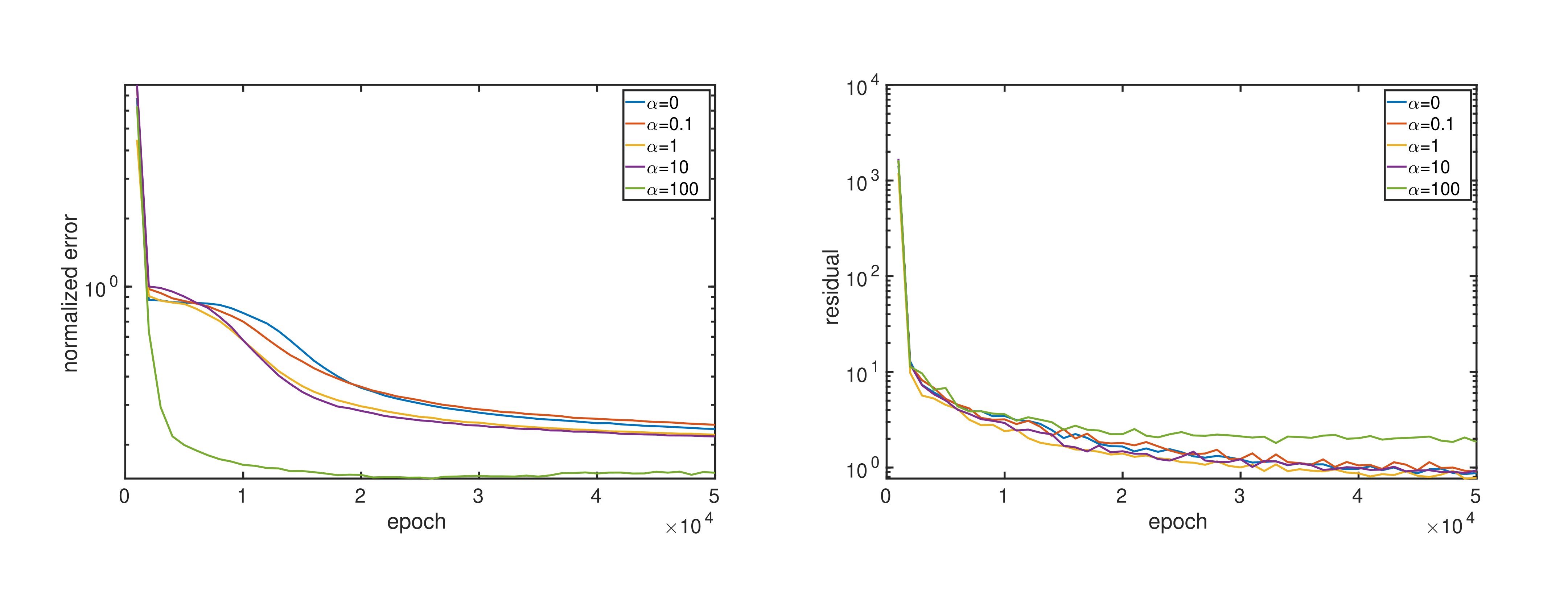}
\end{overpic}
\caption{Comparison of neural network performance for different values of $\alpha$ in the data loss term. \textbf{Left:} epoch vs normalized error given by eq. \ref{eq:relerr} and \textbf{right:} the residual. Both both plots are the done on the test data set.}
\label{fig:neuralnettraining}
\end{figure}
In this section we describe the effect of adding data to the PINN loss function. Given a PINN loss:
\begin{equation}
    f(\varepsilon; \theta) = \bigg|\bigg| \big[ D^h_xD^e_x+D^h_yD^e_x+\omega_o^2 \varepsilon \big] NN(\varepsilon; \theta) +i\omega_o J\bigg|\bigg|_1
    \label{sup:eq:pinn}
\end{equation}
and a data loss
\begin{equation}
g(\varepsilon; \theta) = \bigg|\bigg|E_{FDFD}-E_{NN}(\varepsilon; \theta)\bigg|\bigg|_1
    \label{sup:eq:dataloss}
\end{equation}
we can form a total loss function:
\begin{equation}
h(\varepsilon; \theta) = f(\varepsilon;\theta)+g(\varepsilon;\theta)
    \label{sup:eq:totalloss}
\end{equation}
Where $\varepsilon$ is the input dielectric distribution, $\theta$ are the trainable parameters of the neural network, and $\alpha$ is the "strength" of the data term. Fig. \ref{fig:neuralnettraining} shows the test loss vs epoch of neural networks trained with various $\alpha$ parameters. We generated a dataset of 10000 fields by directly simulating our problem with random pillar arrangements using angler \cite{sref3}. The training was done on 9900 fields, and the test was done on the remaining 100. The left hand side of Fig. \ref{fig:neuralnettraining} shows the accuracy of the neural network given by the 2 norm relative error between the fields simulated by FDFD and the neural net:
\begin{equation}
 \frac{||E_{FDFD}-NN||^2_2}{||E_{FDFD}||_2^2}   
\label{eq:relerr}
\end{equation}
on the test data set. The right hand side shows the same information except for the residual given by \ref{sup:eq:pinn}. We note that, while some data parameter $\alpha$ may marginally improve how well the neural network predicts a field, for values of $\alpha$ where this improvement becomes significant, the physics informed loss takes a penalty. We argue that having a lower physics informed loss is a much better scenario for optimization problems because it means the fields satisfy the PDE better, and thus more accurately represent the predicted physical quantity. Furthermore, the improvement from including a data term is marginal, and increases the complexity of neural network training, that it is not worth adding.
\newpage
\section{Extended depth of focus (EDOF) Lens}
\label{s-edof}
\begin{figure}[t]
\begin{overpic}[width=\textwidth]{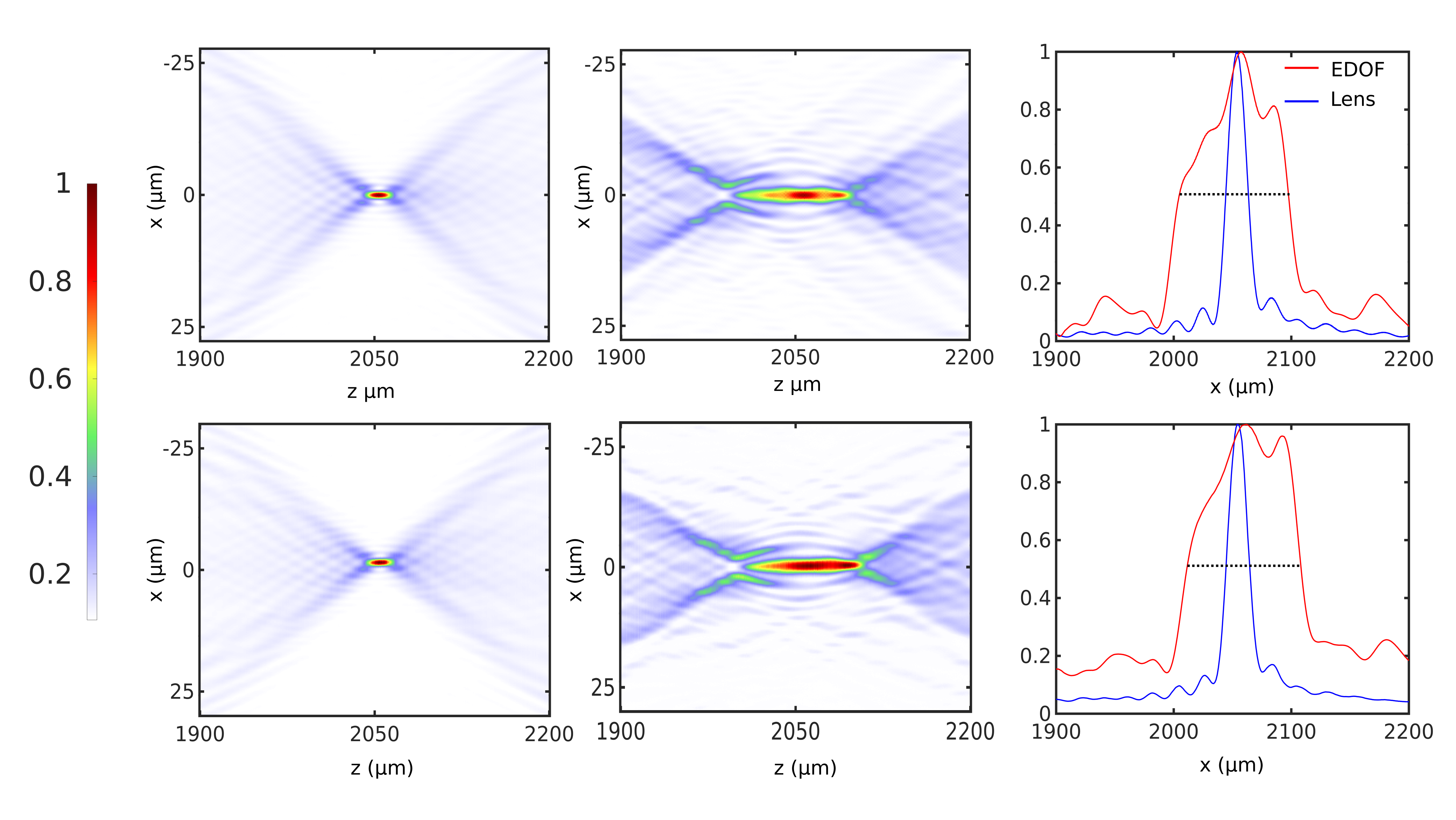}
\end{overpic}
\caption{EDOF lens inverse design. \textbf{a. - c.} represent theoretical results, and \textbf{d. - f.} are experimentally measured results. \textbf{a., d.} are forward designed lenses with focal length 2.05 mm. \textbf{b.,e.} are the optimized EDOF lenses. \textbf{c., f.} are slices along the z axis with $x=0\mu$m. The red lines are the EDOF and the blue lines are the lens. The black line is plotted at the full width half maxima of the EDOF lens, which is how we define the depth of focus\cite{sref5}. The theoretical and experiment depth of focus for the forward designed lens is 20$\mu$m. The EDOF lens has a theoretical depth of focus of 93$\mu$m and an experimentally measured depth of focus of 97$\mu$m}.
\label{fig:edof}
\end{figure}

We also designed an EDOF lens through a standard max-min objective approach, where we computed the intensity of the field produced by our lens at discrete equidistantly spaced points along a focal line, on an interval between two different focal lengths:
\begin{equation}
\begin{array}{c}
     f = \{ f_1, f_2, ..., f_{10} \}  \\
     f_i = E^\dagger(0,z_i) E(0,z_i) \\
     \max\limits_{r} \min\limits_{f_i} f 
\end{array}
\end{equation}
We then used forward design to generate a lens with focal length 2050$\mu$m and a diameter of 1000$\mu$m, and optimized an EDOF lens to extend its focus between 2000 and 2100 microns. Fig. \ref{fig:edof} shows a summary of the results for the inverse designed EDOF lens. To characterize the performance of the lens, we define the depth of focus as the point where the distance at which the focal spot intensity reaches 1/2 of its original value, see Fig. \ref{fig:edof} c and f. The EDOF device has a theoretically predicted depth of focus of 93$\mu$m and an experimentally measured depth of focus of 97$\mu$m. The depth of focus of the forward designed lens is $\sim$20$\mu$m measured both in experiment and theory.
\newpage
\begin{figure}[h!]
\begin{overpic}[width=\textwidth]{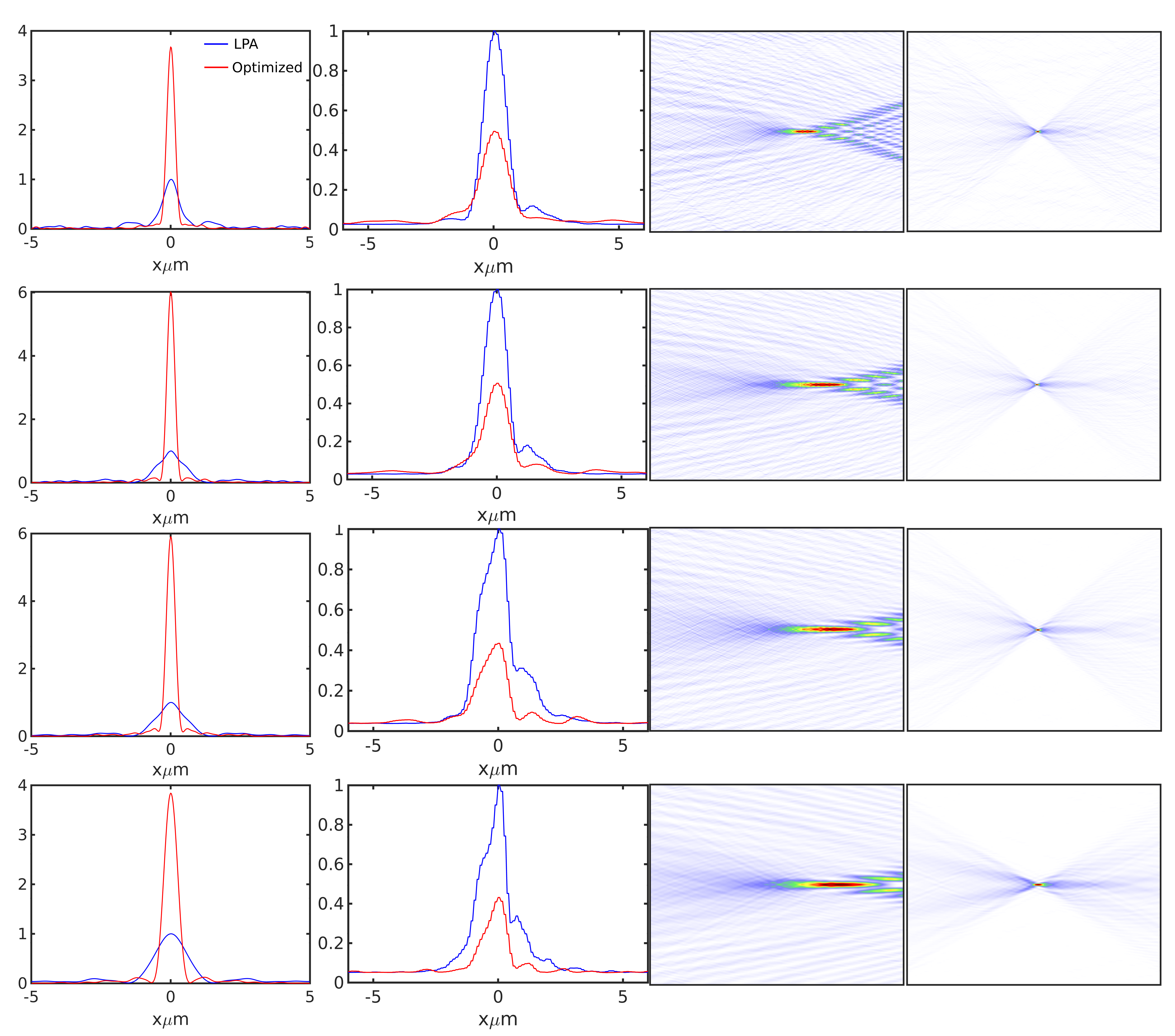}
\put (13,90) {I.}
\put (41,90) {II.}
\put (64,90) {III.}
\put (89,90) {IV.}

\put (-3,79) {a.}
\put (-3,55) {b.}
\put (-3,35) {c.}
\put (-3,11) {d.}
\end{overpic}
\caption{\textbf{I} Theoretically predicted intesnities at the focal spot. \textbf{II.} Experimentally measured intensities at the focal spot. \textbf{III.} Theoretically predicted focal spot for forward design \textbf{IV} and inverse design. \textbf{a.} 200 $\mu$m focal length. \textbf{b.} 500 $\mu$m \textbf{c.} 750 $\mu$m \textbf{d.} 1000$\mu$m.}
\label{fig:failure}
\end{figure}

\newpage

\section{Failed designs and the lack of experimental results}
\label{s-failed-designs}
We want to emphasize the importance of experimentally verifying inverse design methods. Initially we designed our inverse design method by extracting derivative matrices from angler \cite{sref3}, and comparing our results to the FDFD results from the same package in order to make sure our results were consistent and accurate. Fig. \ref{fig:failure} summarizes our results. We note there there is a large discrepancy between the theory and experiment here. This is due to the fact that angler and our first iteration of our inverse design methodology was predicted the complex conjugate of the electric field instead of the electric field, leading to angular spectrum propagation to propagate the fields in the opposite direction. We would never have realized this error if we did not conduct experimental testing of our devices. We encourage more inverse design papers in the future to manufacture and test their inverse design methods in order to produce reliable and accurate designs.
\newpage
\section{Simulation resource and speed comparisons}
\label{s-resources}
\begin{table}[]
\centering
\resizebox{\textwidth}{!}{%
\begin{tabular}{lrrrrrl}
 & \multicolumn{4}{c}{PINN} & \multicolumn{2}{c}{FDFD} \\
 & \multicolumn{1}{l}{Batch Size 1} & \multicolumn{1}{l}{Batch Size 10} & \multicolumn{1}{l}{Batch Size 20} & \multicolumn{1}{l}{Batch size 40} & \multicolumn{1}{l}{Overlapping BCs} & Full Simulation \\
average time per chunk (s) & 0.006 & 0.004 & 0.004 & 0.005 & 0.021 &  \\
total sim time (s) & 1.17 & 1.04 & 1.17 & 1.13 & 5.29 & \multicolumn{1}{r}{11.43} \\
Ram Usage (GB) & 5.29 & 5.29 & 5.29 & 5.29 & 1.77 & \multicolumn{1}{r}{36.7} \\
GPU Memory (GB) & 1.62 & 2.88 & 4.58 & 7.17 & \multicolumn{1}{l}{} & 
\end{tabular}%
}
\caption{Resource and speed comparisons between the PINN approach and the FDFD approach for the forward simulation of a 1mm lens. The PINN approach is about 5x faster than the FDFD approach when overlapping boundary conditions are used and 10x faster when we don't use overlapping boundary conditions. Furthermore, the overlapping boundary method is more memory efficient, and thus useful for running inverse design on machines with low RAM.}
\end{table}

\begin{figure}[h!]
\center
\begin{overpic}[width=\textwidth]{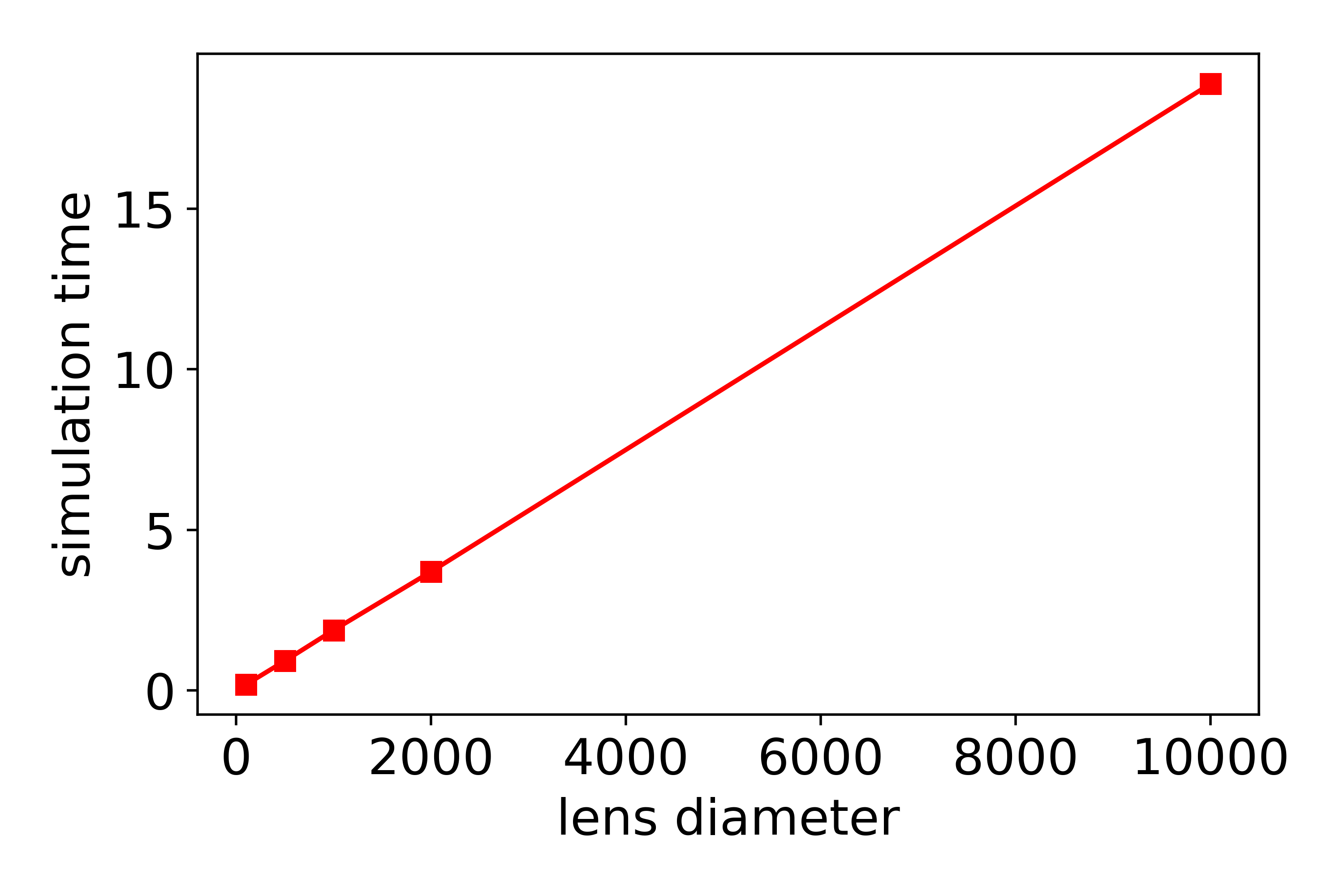}
\end{overpic}
\caption{Comparison of neural network simulation time vs lens diameter. Data is taken for lens diameters of 100, 500, 1000, 2000, and 10000 $\mu$m lens diameters. Simulation time increases linearly with lens diameter.}
\label{fig:neuralnetspeed}
\end{figure}
\newpage
\section{LPA Data}
\begin{figure}[h!]
\center
\begin{overpic}[width=\textwidth]{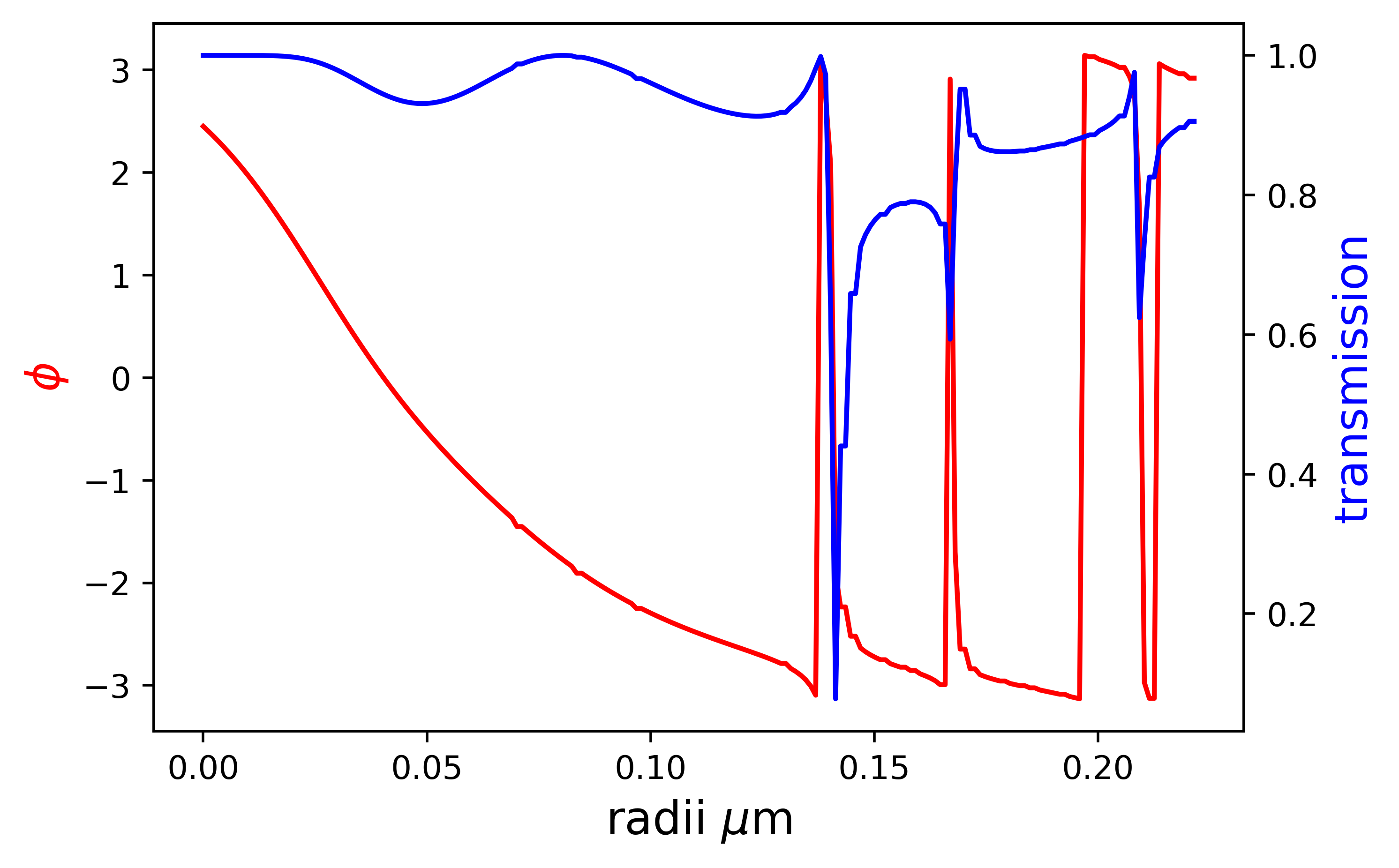}
\end{overpic}
\caption{Phase and amplitude transmission for the pillars with height $h=0.6$, refractive index $n=2$, and periodicity $0.443\mu$m at operating wavelength $\lambda=0.633\mu$m}
\label{fig:neuralnetspeed}
\end{figure}

\end{document}